\documentclass{jetpl}
\twocolumn
\usepackage{epsfig}
\usepackage{graphicx}
\usepackage{amsmath}
\usepackage{bm}
\usepackage{amssymb}
\usepackage{color}

\newif\ifold             \oldtrue            
\def\ba{\begin{eqnarray}}
\def\ea{\end{eqnarray}}
\newcommand{\be}{\begin{equation}}
\newcommand{\ee}{\end{equation}}

\lat

\title{Energy gaps at  neutrality point in bilayer graphene in a magnetic field}

\rtitle{Energy gaps at  neutrality point \ldots}

\sodtitle{Energy gaps at  neutrality point in bilayer graphene in a
magnetic field}

\author{E.\,V.\,Gorbar$^{+}$\/,
V.\,P.\,Gusynin$^+$\/\thanks{e-mail: vgusynin@bitp.kiev.ua},
V.\,A.\,Miransky$^{*}$\/ }

\rauthor{E.\,V.\,Gorbar, V.\,P.\,Gusynin, V.\,A.\,
Miransky  }

\sodauthor{Gorbar,Gusynin,Miransky}

\address{$^+$Bogolyubov Institute for Theoretical Physics, 03680, Kiev, Ukraine\\~\\
$^*$Department of Applied Mathematics, University of Western
Ontario, London, Ontario N6A 5B7, Canada}

\dates{15 January 2010}{*}

\abstract{Utilizing the Baym-Kadanoff formalism with the
polarization function calculated in the random phase approximation,
the dynamics of the $\nu=0$ quantum Hall state in bilayer graphene
is analyzed. Two phases with nonzero energy gap, the ferromagnetic
and layer asymmetric ones, are found. The phase diagram in the plane
$(\tilde{\Delta}_0,B)$, where $\tilde{\Delta}_0$ is a top-bottom
gates voltage imbalance, is described. It is shown that the energy
gap scales linearly, $\Delta E\sim 14 B[{\mbox T}]{\mbox K}$, with
magnetic field.}


\begin{document}

\maketitle

%\pacs{81.05.ue, 73.43.-f, 73.43.Cd}

%%%PACS from 2010
%%% 81.05.ue
%%% 73.43.-f
%%% 73.43.Cd

{\em Introduction.---}
%The properties of bilayer graphene
%\cite{McC,Nov,CN}, consisting of two closely coupled graphene layers, have
%attracted great interest.
The possibility of inducing and controlling the
energy gap by gates voltage makes bilayer graphene \cite{McC,Nov,CN}
one of the most active
research areas with very promising  applications in electronic devices. Recent
experiments in bilayer graphene \cite{footn1,Zh} showed the generation of gaps
in a magnetic field with complete lifting of the eight-fold degeneracy in the
zero energy Landau level, which leads to new quantum Hall states with filling
factors $\nu=0,\pm1,\pm2,\pm3$. Besides that, in suspended bilayer graphene,
Ref.\cite{footn1} reports the observation of an extremely large
magnetoresistance in the $\nu=0$ state due to the energy gap $\Delta E$, which
scales linearly with a magnetic field $B$, $\Delta E \sim 3.5-10.5
B[{\mbox T}]{\mbox K}$, for $B \lesssim 10{\mbox T}$. This linear scaling is
hard to explain by the standard  mechanisms \cite{QHF,MC} of gap generation
used in a monolayer graphene, which lead to large gaps of the order of the
Coulomb energy $e^{2}/l \sim B^{1/2}$, $l=(\hbar c/eB)^{1/2}$ is the magnetic
length.

In this Letter, we study the dynamics of clean bilayer graphene in a
magnetic field, with the emphasis on the $\nu=0$ state in the
quantum Hall effect (QHE). It will be shown that, as in the case of
monolayer graphene \cite{Gorb1}, the dynamics in the QHE in bilayer
graphene is described by the {\em coexisting} quantum Hall
ferromagnetism (QHF) \cite{QHF} and magnetic catalysis (MC)\cite{MC}
order parameters. The essence of the dynamics is an effective
reduction by two units of the spatial dimension in the electron-hole
pairing in the lowest Landau level (LLL) with energy $E=0$
\cite{catal,Khvesh,Gorb}. As we discuss below, there is however an
essential difference between the QHE's in these two systems. While
the pairing forces in monolayer graphene lead to a relativistic-like
scaling $\Delta E \sim \sqrt{|eB|}$ for the dynamical gap, in
bilayer graphene, such a scaling takes place only for strong
magnetic fields, $B \gtrsim B_{thr}$, where our estimate yields
$B_{thr} \sim 30-60{\mbox T}$. For $B \lesssim B_{thr}$, a
nonrelativistic-like scaling $\Delta E\sim |eB|$ is realized in the
bilayer. The origin of this phenomenon is very different forms of
the polarization function in monolayer graphene and bilayer one that
in turn is determined by the different dispersion relations for
quasiparticles in these two systems. The polarization function is
one of the major players in the QHE in bilayer, and its
consideration distinguishes this work from the most of previous
theoretical ones studying the QHE in bilayer graphene \cite{theory}
\footnote{The polarization effects in bilayer graphene were recently
considered in \cite{NL}, however, the authors used a polarization
function with no magnetic field for their estimate.}.

Using the random phase approximation in the analysis of the gap
equation, we found that the gap in the clean bilayer is $\Delta E\sim
14 B[{\mbox T}] {\mbox K}$ for the magnetic field $B \lesssim B_{thr}$. The
phase diagram in the plane $(\tilde{\Delta}_0,B)$, where $\tilde{\Delta}_0$ is
a top-bottom gates voltage imbalance, is described. These
are the central results of this Letter.

{\em Hamiltonian.---} The free part of the effective low energy Hamiltonian of
bilayer graphene is \cite{McC}:
\be
H_0 = - \frac{1}{2m}\int
d^2x\Psi_{Vs}^+(x)\left( \begin{array}{cc} 0 & (\pi^{\dagger})^2\\ \pi^2 & 0
\end{array} \right)\Psi_{Vs}(x), \label{free-Hamiltonian}
\ee
 where $\pi=\hat{p}_{x_{1}}+i\hat{p}_{x_{2}}$ and the canonical momentum
$\hat{\mathbf{p}} = -i\hbar\bm{\nabla}+ {e\mathbf{A}}/c$ includes the
vector potential $\mathbf{A}$ corresponding to the external magnetic field
$\mathbf{B}$. Without magnetic field, this Hamiltonian generates the spectrum
$E=\pm \frac{p^2}{2m}$, $m= \gamma_1/2v_{F}^2$, where the Fermi velocity $v_F
\simeq c/300$ and $\gamma_1 \approx 0.34-0.40$eV. The two component spinor
field $\Psi_{Vs}$ carries the valley $(V= K, K^{\prime})$ and spin
$(s = +, -)$ indices. We will use the standard convention:
$\Psi_{Ks}^T=(\psi_A{_1}, \psi_B{_2})_{Ks}$
whereas $\Psi_{K^{\prime}s}^T = (\psi_B{_2},
\psi_A{_1})_{K^{\prime}s}$. Here $A_1$ and $B_2$ correspond to
those sublattices in the layers 1 and 2, respectively, which, according
to Bernal $(A_2-B_1)$ stacking, are relevant for the low energy dynamics.
The effective Hamiltonian (\ref{free-Hamiltonian}) is valid for magnetic fields
$1T < B < B_{thr}$. For $B < 1T$, the trigonal warping should be taken into
account \cite{McC}. For $B> B_{thr}$, a monolayer like Hamiltonian with linear
dispersion should be used.

The Zeeman and Coulomb interactions in bilayer graphene are
(henceforth we will omit indices $V$ and $s$ in the field $\Psi_{Vs}$):
\ba
&&\hspace{-5mm}H_{int}= \mu_{B}B\hspace{-1.5mm} \int\hspace{-1.0mm} d^2x \Psi^+(x)\sigma^3\Psi(x) +
\frac{e^2}{2\kappa}\int\hspace{-1.0mm}
d^3xd^3x^{\prime}\frac{n(\mathbf{x})n(\mathbf{x}^{\prime})}{|\mathbf{x}-
\mathbf{x}^{\prime}|}\nonumber\\
&&\hspace{-3.5mm}= \mu_{B}B\hspace{-1.0mm} \int\hspace{-1.0mm} d^2x \Psi^+(x) \sigma^3\Psi(x) +
\frac{1}{2}\int \hspace{-1.0mm}d^2xd^2x^{\prime}\left[
V(x-x^{\prime})\right.\nonumber\\
&&\hspace{-5mm}\left.\times\left(\rho_1(x)\rho_1(x^{\prime})+
\rho_2(x)\rho_2(x^{\prime})\right)\hspace{-1.0mm} +
2V_{12}(x-x^{\prime})\rho_1(x)\rho_2(x^{\prime}) \right],\nonumber\\
\label{interaction}
\ea
where $\mu_B$ is the Bohr magneton, $\kappa$ is the
dielectric constant, and
$n(\mathbf{x})=\delta(z-\frac{d}{2})\rho_1(x)+\delta(z+\frac{d}{2})\rho_2(x)$
is the three dimensional charge density ($d \simeq 0.3$nm is the distance
between the two layers). The interaction potentials $V (x)$ and $V_{12}(x)$
describe the intralayer and interlayer interactions, respectively.
Their Fourier transforms are ${V}(k)= 2\pi e^2/\kappa k$ and
${V}_{12}(k)= 2\pi e^2e^{-kd}/\kappa k$. The two-dimensional charge
densities $\rho_1(x)$ and $\rho_2(x)$ are:
\be
\rho_1(x)=\Psi^+(x)P_1\Psi(x)\,,\quad \rho_2(x)=\Psi^+(x)P_2\Psi(x)\,,
\label{density}
\ee
where $P_1=\frac{1+\xi\tau^3}{2}$ and
$P_2=\frac{1-\xi\tau^3}{2}$ are projectors on states in the layers 1 and 2,
respectively [here $\tau^3$ is the Pauli matrix acting on layer components, and
$\xi = \pm 1$ for the valleys $K$ and $K^{\prime}$, respectively].

{\em Symmetries.---} The Hamiltonian $H = H_0 + H_{int}$ describes the dynamics
at the neutral point (with no doping). Because of the projectors $P_1$ and
$P_2$ in charge densities (\ref{density}), the symmetry of the Hamiltonian $H$
is essentially lower than the symmetry in monolayer graphene. If the Zeeman
term is ignored, it is $U^{(K)}(2)_S \times U^{(K^{\prime})}(2)_S\times
Z_{2V}^{(+)}\times Z_{2V}^{(-)}$, where $U^{(V)}(2)_S$ defines the $U(2)$ spin
transformations in a fixed valley $V = K, K^{\prime}$, and $Z_{2V}^{(s)}$
describes the valley transformation $\xi \to -\xi$ for a fixed spin $s = \pm$
(recall that in monolayer graphene the symmetry would be $U(4)$ \cite{Gorb}).
The Zeeman interaction lowers this symmetry down to $G_2 \equiv U^{(K)}(1)_{+}
\times U^{(K)}(1)_{-} \times U^{(K^{\prime)} }(1)_{+} \times U^{(K^{\prime)}
}(1)_{-} \times Z_{2V}^{(+)}\times Z_{2V}^{(-)}$, where $U^{(V)}(1)_{s}$ is the
$U(1)$ transformation for fixed values of both valley and spin. Recall that the
corresponding symmetry in monolayer graphene is $G_1 \equiv U^{(+)}(2)_V \times
U^{(-)}(2)_V$, where $U^{(s)}(2)_V$ is the $U(2)$ valley transformations for a
fixed spin.

{\em Order parameters.---} Although the $G_1$ and $G_2$ symmetries are quite
different, it is noticeable that their breakdowns can be described by the same
QHF and MC order parameters. The point is that these $G_1$ and $G_2$ define the
same four conserved commuting currents whose charge densities (and four
corresponding chemical potentials) span the QHF order parameters (we use the
 notations of Ref. \cite{Gorb1}):
\begin{eqnarray}
\label{mu}
\mu_s:\quad
\Psi^\dagger_s\Psi_s &=& \psi_{K  A_1 s}^\dagger\psi_{K A_1 s}
+\psi_{K^{\prime} A_1 s}^\dagger\psi_{K^{\prime}A_1 s}\nonumber\\
& +&\psi_{K B_2 s}^\dagger\psi_{K B_2 s}
+ \psi_{K^{\prime}B_2 s}^\dagger\psi_{K^{\prime} B_2 s}\,,\\
\label{mu-tilde}
\tilde{\mu}_{s}:\quad
\Psi^\dagger_s\xi\Psi_s &=& \psi_{K  A_1 s}^\dagger\psi_{K A_1 s}
- \psi_{K^{\prime} A_1 s}^\dagger\psi_{K^{\prime}A_1 s}\nonumber\\
& +& \psi_{K B_2 s}^\dagger\psi_{K
B_2 s} - \psi_{K^{\prime}B_2 s}^\dagger\psi_{K^{\prime} B_2 s}\,.
\end{eqnarray}
The order parameter (\ref{mu}) is the charge density for a fixed spin
whereas the order parameter (\ref{mu-tilde}) determines the charge-density
imbalance between the two valleys. 
The corresponding chemical potentials are
$\mu_s$ and $\tilde{\mu}_{s}$, respectively.
While the former order 
parameter preserves the $G_2$ symmetry, the latter completely
breaks its discrete subgroup $Z_{2V}^{(s)}$.
Their MC cousins are
\begin{eqnarray}
\label{delta}
\Delta_s: \quad
\Psi^\dagger_s\tau_3\Psi_s &=& \psi_{K  A_1 s}^\dagger\psi_{K A_1 s}
-\psi_{K^{\prime} A_1 s}^\dagger\psi_{K^{\prime}A_1 s}\nonumber\\
 &-&\psi_{K B_2 s}^\dagger\psi_{K
B_2 s}+ \psi_{K^{\prime}B_2 s}^\dagger\psi_{K^{\prime} B_2 s}\,,\\
\label{delta-tilde}
\tilde{\Delta}_{s}:\quad
\Psi^\dagger_s\xi\tau_3\Psi_s&=& \psi_{K  A_1 s}^\dagger\psi_{K A_1 s}
+\psi_{K^{\prime} A_1 s}^\dagger\psi_{K^{\prime}A_1 s}\nonumber\\
& -&\psi_{K B_2 s}^\dagger\psi_{K
B_2 s} - \psi_{K^{\prime}B_2 s}^\dagger\psi_{K^{\prime} B_2 s}\,.
\end{eqnarray}
These order parameters can be rewritten in the form of Dirac mass terms \cite{Gorb1} corresponding to the masses $\Delta_s$ and 
$\tilde{\Delta}_{s}$, respectively.
While the order parameter (\ref{delta})
preserves the $G_2$, it is odd under time reversal $\cal{T}$
\cite{Hald}. On the other hand, the order parameter (\ref{delta-tilde}) is connected
with the conventional Dirac mass $\tilde{\Delta}$. It determines the charge-density
imbalance between the two layers \cite{McC}. Like
$\tilde{\mu}_s$, this mass term completely breaks the $Z_{2V}^{(s)}$
symmetry and is even under $\cal{T}$. Note that because of the
Zeeman interaction, the $SU^{(V)}(2)_S$ is explicitly broken, leading to a
spin gap. This gap could be dynamically strongly enhanced \cite{Aban}. In that
case, a quasispontaneous breakdown of the $SU^{(V)}(2)_S$ takes place.
The corresponding ferromagnetic phase is described by
$\mu_3 = (\mu_+ - \mu_-)/2$
with the QHF order parameter $\Psi^\dagger\sigma_3\Psi$, and by
$\Delta_3 = (\Delta_+ - \Delta_-)/2$ with the MC order parameter
$\Psi^\dagger\tau_3\sigma_3\Psi$ \cite{Gorb1}.

{\em Gap equation.---} In the framework of the Baym-Kadanoff formalism
\cite{BK}, and using the polarization function calculated in the random phase
approximation (RPA), we analyzed the gap equation for the LLL quasiparticle
propagator with the order parameters introduced above. Recall that in bilayer
graphene, the LLL includes both the $n=0$ and $n=1$ LLs, if the Coulomb
interaction is ignored \cite{McC}. Therefore there are sixteen parameters
$\mu_{s}(n)$, $\Delta_{s}(n)$, $\tilde{\mu}_{s}(n)$, and
$\tilde{\Delta}_{s}(n)$, where the index $n= 0, 1$ corresponds to the $n=0$ and
$n=1$ LLs, respectively. The following system of equations was derived for
these parameters:
\ba
G^{-1}_{\xi s0}(\Omega)&=&S^{-1}_{\xi s}(\Omega)-
i\int\frac{d\omega\,d^{2}k}{(2\pi)^{3}}\,e^{-\mathbf{k}^{2}l^{2}/2}
[G_{\xi s0}(\omega)\nonumber\\
&+& G_{\xi s1}(\omega)\mathbf{k}^{2}l^{2}/2]
{V}_{eff}\left(
\Omega-\omega,|\mathbf{k}|\right)\nonumber\\
&-&\frac{e^2 d}{2\kappa l^2}\,\left(\frac{1+\xi}{2}A_1 +\frac{1-\xi}{2}A_2\,
\right),
\label{gap-equation-g-2}\\
G^{-1}_{\xi s1}(\Omega)&=&S^{-1}_{\xi
s}(\Omega)-i\int\frac{d\omega\,d^{2}k}{(2\pi)^{3}}\,
e^{-\mathbf{k}^{2}l^{2}/2}[G_{\xi s0}(\omega)\nonumber\\
&\times&\mathbf{k}^{2}l^{2}/2+G_{\xi s1}(\omega)(1-\mathbf{k}^{2}l^{2}/2)^{2}]
 \nonumber\\
&\times&V_{eff}\left( \Omega-\omega,|\mathbf{k}|\right)\nonumber\\
&-&\frac{e^2
d}{2\kappa l^2}\, \left(\frac{1+\xi}{2}A_1 +\frac{1-\xi}{2}A_2
\right)\,. \label{gap-equation-g-3}
\ea
Here  $A_1 =
\sum_{n,s}\,\mbox{sgn}(\,E_{- n s })$\,, $A_2 =
\sum_{n,s}\,\mbox{sgn}(\,E_{+ n s})$, and
\be
 S_{\xi s}(\omega)=\frac{1}{\omega+ \mu_{0} - sZ +\xi\tilde{\Delta}_{0}},\,
 G_{\xi s n}(\omega)=\frac{1}{\omega-E_{\xi  n s}}
\ee
are frequency dependent factors in the bare and full LLL propagators, where
\be E_{\xi n s}= -(\mu_{s}(n)+ \Delta_{s}(n))+ \xi(\tilde{\mu}_{s}(n) -
\tilde{\Delta}_{s}(n)) \label{disp}
\ee
are the energies of the LLL states,
$\mu_{0}$ is chemical potential, $Z$ is the Zeeman energy,
$Z\simeq\mu_{B}B=0.67\,B[{\mbox T}]{\mbox K}$.
%, $\tilde{\Delta}_{0}$ is a top-bottom gates voltage imbalance.
The second and third terms on
right hand sides of Eqs.(\ref{gap-equation-g-2}), (\ref{gap-equation-g-3})
describe the Fock and Hartree interactions, respectively. Note that because
for the LLL states only the component $\psi_{B_2s}$ $(\psi_{A_1s})$ of the wave
function at the $K (K^{\prime})$ valley is nonzero, their energies depend only
on the eight independent combinations of the QHF and MC parameters shown in
Eq.(\ref{disp}).
The function ${V}_{eff}(\omega,k)$, describing the Coulomb interaction,
is
\be
{V}_{eff}(\omega,k)=\frac{2\pi e^2}{\kappa}\,\frac{1}{ k
+\frac{4\pi e^2}{\kappa}\Pi (\omega,{\bf k}^2)}, \label{V-D}
\ee
where $\Pi(\omega,{\bf k}^2)$ is the polarization function in a magnetic field. Since the
dependence of $\Pi (\omega,{\bf k}^2)$ on $\omega$ is weak, the static
polarization will be used. Then, in the case of frequency independent order
parameters, the integration over $\omega$ in Eqs. (\ref{gap-equation-g-2}),
(\ref{gap-equation-g-3}) can be performed explicitly, and we get a system of
algebraic equations for the energies $E_{\xi ns}$ of the LLL states.

\begin{figure}[htp]
\includegraphics[width=7.0cm]{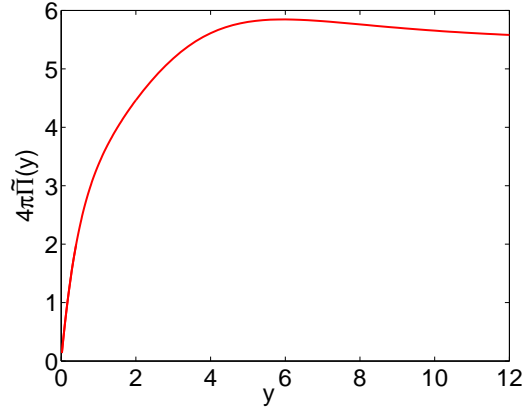}
\caption{Fig. 1. The static polarization function $4\pi\tilde{\Pi}(y)$. }
\label{fig1}
\end{figure}

It is convenient to rewrite the static polarization $\Pi(0,{\bf
k}^2)$ in the form $\Pi = (m/{\hbar}^2)\tilde{\Pi}(y)$, where both
$\tilde{\Pi}$ and $y \equiv \mathbf{k}^2l^2/2$ are dimensionless.
The function $\tilde{\Pi}(y)$ was expressed in terms of the sum over
all the Landau levels and was analyzed both analytically and
numerically. 
At $y\ll 1$, it behaves as $\tilde{\Pi}(y) \simeq 0.55y$ and its
derivative $\tilde{\Pi}'$ changes from $0.55$ at $y=0$ to 0.12 at
$y=1$. At large $y$, it approaches a zero magnetic field value,
$\tilde{\Pi}(y)\simeq \ln4/\pi$ (see Fig.1)\,\footnote{ One
can show that the presence of a maximum in the function $4\pi
\tilde{\Pi}(y)$ in Fig. \ref{fig1} follows from the equality of the
polarization charge density $n(r)$ in a magnetic field $B$ and that
at $B=0$ as $r \to 0$.}.
Because of the Gaussian factors $e^{-\mathbf{k}^{2}l^{2}/2} = e^{-y}$ in Eqs.
(\ref{gap-equation-g-2}) and (\ref{gap-equation-g-3}), the relevant region in
the integrals in these equations is $0 < y \lesssim 1$.
The crucial point in the analysis
is that the region where the bare Coulomb term $k$ in the
denominator of ${V}_{eff}(k)\equiv {V}_{eff}(0,k)$ (\ref{V-D}) dominates is
very small, $0 < y \lesssim 10^{-3}B$[T]. The main reason of that is a large
mass $m$ of quasiparticles, $m \sim 10^{-2}m_e \sim 10^8 {\mbox K/c^2}$.
As a result, the polarization function term dominates in
${V}_{eff}(k)$ that leads to ${V}_{eff}(k)= C(y)\hbar^{2}/ml^2k^{2}$,
where the part with the factor $1/k^{2}$ corresponds
to the Coulomb
potential in two dimensions, and the function $C(y)$ describes its
smooth modulations at $0\leq y \lesssim 1$ (see Fig.1).
It is unlike the case of the monolayer graphene where the effective
interaction is proportional to $1/k$.
As we discuss below, this in turn implies that, in the
low energy model described by the Hamiltonian in Eqs. (\ref{free-Hamiltonian}),
(\ref{interaction}), the scaling $\Delta E \sim |eB|$ takes place for the
dynamical energy gap, and not $\Delta E \sim \sqrt{|eB|}$ taking place in
monolayer graphene \cite{QHF,MC,Gorb1}.

Last but not least, using the model with four-component wave functions
\cite{McC}, we determined the upper limit for the values of $B$, $B_{thr}$, for
which the low energy effective model can be used. We found that $B_{thr} \sim
30-60$T, corresponding to the experimental values $0.34-0.40$eV of the
parameter $\gamma_1 = 2m v^{2}_F$. We predict that for the values $B >
B_{thr}$, the monolayer like scaling, $\Delta E \sim \sqrt{|eB|}$, should take
place.

{\em Solutions.---} At the neutral point ($\mu_0 = 0$, no doping),
we found two competing solutions of Eqs. (\ref{gap-equation-g-2})
and (\ref{gap-equation-g-3}): I) a ferromagnetic (spin splitting)
solution, and II) a layer asymmetric solution, actively discussed in
the literature. The energy (\ref{disp}) of the LLL states of the
solution I equals:
\be
E^{(I)}_{\xi ns }=s(Z+\frac{I_{n}(B)}{2ml^2})
- \xi\tilde{\Delta}_0 \,, \label{solution-I}
\ee
where the notation $I_{n}(B)$ is used for the integrals
\be
I_{0}(B)=\hspace{-0.5mm}
\int\limits_{0}^{\infty}\hspace{-0.5mm}\frac{d
y\,(1+y)e^{-y}}{\sqrt{x y}+ 4\pi \tilde{\Pi}(y)},
I_{1}(B)=\hspace{-0.5mm}\int\limits_{0}^{\infty}\hspace{-0.5mm}\frac{d
y\,(1-y+y^{2})e^{-y}}{\sqrt{x y}+ 4\pi\tilde{\Pi}(y)}
\label{integralsIi}
\ee
with $x=0.003 B(T)$. Note that the Hartree
interaction does not contribute to this solution. The situation is
different for the solution II: \ba
 E^{(II)}_{\xi ns}&=&s Z - \xi(\tilde{\Delta}_0 +
\frac{I_{n}(B)}{2ml^2} - \frac{2e^2 d}{\kappa l^2})\,. \label{solution-II}
\ea
The last term in the parenthesis is the Hartree one. For suspended bilayer
graphene, we will take $\kappa = 1$.

The energy density of the ground state
for these solutions is ($a = I, II$):
\ba
\epsilon^{(a)}=&-&\frac{1}{8\pi l^2}
\sum_{\xi=\pm}\sum_{s=\pm}\sum_{n=0,1}\left[|E_{\xi ns}^{(a)}|\right.\nonumber\\
 &+&\left. (-s\,0.67B+
\xi\tilde{\Delta}_0)\,\mbox{sgn}\,E_{\xi ns}^{(a)}\right]. \label{dens}
\ea
It is easy to check that for balanced bilayer ($\tilde{\Delta}_0 = 0$)
the solution I is favorite. The main reason of this is the presence of the
capacitor like Hartree contribution in the energy density of the solution II:
it makes that solution less stable.
\begin{figure}[htp]
\includegraphics[width=7.0cm]{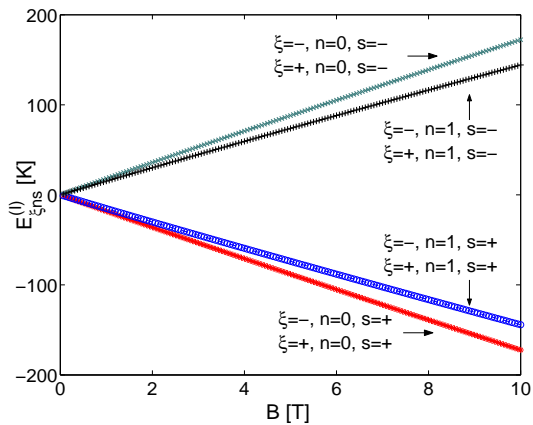}
\caption{Fig. 2. The energies of the LLL states as functions of $B$.
 }
 \label{fig2}
\end{figure}
For $\tilde{\Delta}_0 = 0$, the dependence of the LLL energies $E^{(I)}_{\xi
n s}$ of the solution I on $B$ is shown in Fig.~\ref{fig2} (energy gaps are
degenerate in $\xi$). The perfectly linear form of this dependence is evident.
Also, the degeneracy between the states of the $n=0$ LL and those of the $n =
1$ LL is removed. The energy gap corresponding to the $\nu = 0$ plateau is
$\Delta E = (E_{\xi1 - }^{(I)} - E_{\xi1 + }^{(I)})/2 \simeq 14.3 B [{\mbox
T}]$K.
\begin{figure}[htp]
\includegraphics[width=7.0cm]{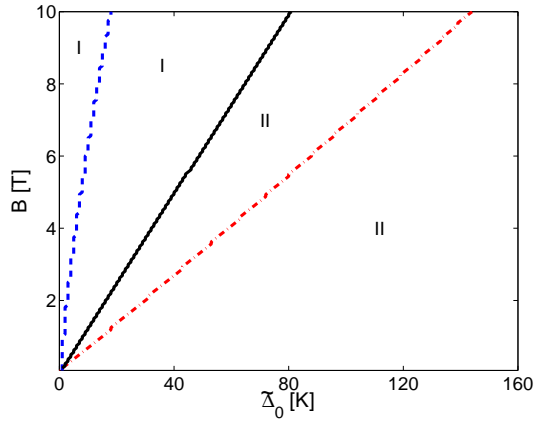}
\caption{Fig. 3.  The phase diagram in the $(\tilde{\Delta}_0,B)$ plane.}
\label{fig3}
\end{figure}
In Fig.~\ref{fig3}, the phase diagram in the plane $(\tilde{\Delta}_0,B)$ is
presented. The  area marked by I (II) is that where the solution I (solution II)
is favorite. The two dashed lines compose the boundary of the region where the
two solutions coexist (the solution I does not exist to the right of the
dashed line in the region II, while the solution II does not exist to the left of the
dashed line in the region I). The  bold line is the line of the first order phase transition.
It is noticeable that for any fixed value of $B\, (\tilde{\Delta}_0)$,
there are sufficiently large values of $\tilde{\Delta}_0$\, (B),
at which the solution I (solution II) does not exist at all. It is because a
voltage imbalance (Zeeman term) tends to destroy the solution I (solution II).

In conclusion, the dynamics of bilayer graphene in a magnetic field
$B \lesssim B_{thr}$ is characterized by a very strong screening of the Coulomb
interaction that relates to the presence of a large mass $m$ in
the nonrelativistic-like dispersion relation for quasiparticles.
The functional dependence of the
gap on $B$ in Fig.~\ref{fig2} agrees with that obtained very recently in
experiments in
Ref. \cite{footn1}. The existence of the first order
phase transition in the plane $(\tilde{\Delta}_0,B)$ is predicted. We also
estimate the value $B_{thr}$, at which the change of the scaling
$\Delta E \sim |eB|$ to $\Delta E \sim \sqrt{|eB|}$ occurs, as
$B_{thr} \sim 30-60$T.
It would be interesting to extend this analysis to the case of the higher,
$\nu = 1, 2,$ and 3, LLL plateaus \cite{footn1,Zh}.

We  thank Junji Jia and S.G. Sharapov for fruitful discussions. The
work of E.V.G and V.P.G. was supported partially by the SCOPES grant
\# IZ73Z0\verb|_|128026 of the Swiss NSF, by the grant SIMTECH
\# 246937 of the European FP7 program, and by the grant RFFR-DFFD \# 28.2/083.
The work of V.A.M. was supported
by  the Natural Sciences and Engineering Research Council of Canada.

\end{document}